\documentclass[twocolumn,showpacs]{revtex4}
\usepackage{graphicx}
\usepackage{dcolumn}
\usepackage{bm}
\usepackage{amsmath}
\usepackage{amsfonts}
\usepackage{amssymb}
\providecommand{\U}[1]{\protect\rule{.1in}{.1in}}

\begin{document}
\title{Decoherence without classicality in the resonant quantum kicked rotor }
\author{A. Romanelli}
\altaffiliation[{\textit{E-mail address:}}\\]{ alejo@fing.edu.uy}
\affiliation{Instituto de F\'{\i}sica, Facultad de Ingenier\'{\i}a\\
Universidad de la Rep\'ublica\\
C.C. 30, C.P. 11000, Montevideo, Uruguay}
\date{\today }
\begin{abstract}
%\vspace{0.2cm} \\
We study the quantum kicked rotor in resonance subjected to an
unitary noise defined through Kraus operators. We show that this
type of decoherence does not, in general, lead to the classical
diffusive behavior. We find exact analytical expressions for the
density matrix and the variance in the primary resonances. The
variance does not loose its ballistic behavior, however the
coherence decays as a power law. The secondary resonances are
treated numerically, obtaining a power-law decay for the variance
and an exponential law decay for the coherence.
\end{abstract}
 \maketitle
\section{Introduction}
The development of experimental techniques has made possible to trap
samples of atoms using resonant exchanges of momentum and energy
between atoms and laser light \cite{cohen}. This progress has been
accompanied by the development of the interdisciplinary fields of
quantum computation and quantum information \cite{Chuang}.

The study of open quantum systems is an outstanding topic of quantum
mechanics. In particular the transition form the quantum to the
classical world has intrinsic importance. On the other hand the
advent of quantum computation makes of decoherence a central problem
in the interaction of the quantum computer with its surroundings.

Simple theoretical and experimental models such as the quantum
kicked rotor (QKR) and the quantum walk (QW) may play an important
role in this frame. Although the existent experiments have high
accuracy in both coherent storage and manipulation of the atoms, the
interaction with the surroundings introduces different degrees of
decoherence influencing the unitary evolution of the system.

The QKR is a milestone in the study of chaos at the quantum level
\cite{CCI79}. The behavior of the QKR depends on whether the period
of the kick is a rational or irrational multiple of $2\pi$ (in
convenient units) \cite{Izrailev}. For rational multiples the
behavior of the system is resonant with ballistic spreading  and has
no classical analog, its standard deviation $\sigma$ has the time
dependence $\sigma(t)\sim t$. For irrational multiples the average
energy of the system grows in a diffusive manner for a short time
and then dynamical localization take place. The quantum resonances
and the dynamical localization of the QKR have been experimentally
observed in samples of cold atoms interacting with a far-detuned
standing wave of laser light \cite{Moore} and in particular the
secondary resonances have been recently observed by Kanem et al.
\cite{Kanem}.

The QKR as a simple toy model allows to study the complexity of
decoherence both analytically and numerically, these studies have a
$25$ year-old history \cite{Ott,Dittrich1,Dittrich2}. On the other
hand the first experimental observation of environment induced
decoherence in the QKR was reported by Ammann et \emph{al.}
\cite{Ammann}.

In this line we recently investigated the QKR in resonance subjected
to: a) decoherence with a L\'evy waiting-time distribution
\cite{alejo1,alejo2} and b) an excitation that follows an aperiodic
Fibonacci prescription \cite{alejo3}. In both cases we find that the
secondary resonances have a sub-ballistic behavior ($\sigma(t)\sim
t^{c},1/2<c<1$), while the principal resonances maintain the
well-known ballistic behavior. These results are very surprising
since one expects diffusive behavior when decoherence occurs. Other
authors also investigated the QKR subjected to noises with a L\'evy
distribution \cite{Schomerus1,Schomerus2} and almost periodic
Fibonacci sequence \cite{Casati0}, showing that this decoherence
never fully destroys the dynamical localization of the system, but
leads to a sub-diffusion regime for a short time before localization
appears.

In this work we want to study the decoherence effect of an unitary
operation described by Kraus operators \cite{Kraus} acting on the
density matrix. Our route is similar to that followed by Brun
\emph{et. al} \cite{Brun} with the QW but our results in the QKR are
very different.
\section{Kicked rotor}
In this section we briefly review the dynamical equations for the
QKR \cite{Izrailev}. Its Hamiltonian is
\begin{equation}
H=\frac{P^{2}}{2I}+K\cos\theta\sum_{n=1}^{\infty}\delta(t-nT)
\label{qkr_ham}
\end{equation}
where the external kicks occur at times $t=nT$ with $n$ integer and
$T$ the kick period, $I$ is the moment of inertia of the rotor, $P$
the angular momentum operator, $K$ the strength parameter and
$\theta$ the angular position. In the angular momentum
representation, $P|\ell\rangle=\ell \hbar|\ell\rangle$, the matrix
element of the time-step evolution operator $U$ is
\begin{equation}
U_{\ell j}\equiv\left\langle \ell\right| U(\kappa)\left|
j\right\rangle=i^{-(j-\ell)}e^{-ij^{2}\varepsilon
T/\hbar}\,J_{j-\ell}(\kappa), \label{evolu}
\end{equation}
where $\varepsilon=\hbar^{2}/2I$, $J_{m}$ is the $m$th order
cylindrical Bessel function and its argument is the dimensionless
kick strength $\kappa\equiv K/\hbar$. The resonance condition does
not depend on $\kappa$ and takes place when the frequency of the
driving force is commensurable with the frequencies of the free
rotor. Inspection of eq.(\ref{evolu}) shows that the resonant values
of the scale parameter $\tau\equiv\varepsilon T/\hbar$ are the set
of the rational multiples of $2\pi$, $\tau=2\pi$ $p/q$. In what
follows we assume, that the resonance condition is satisfied,
therefore the evolution operator depends on $\kappa$, $p$ and $q$.
We call a resonance primary when $p/q$ is an integer and secondary
when it is not.

\section{Kicked rotor dynamics with decoherence}

In order to generate the dynamics of the system we consider that the
decoherence is introduced through a completely positive map, that is
defined by a set of Kraus operators $\{{A_{n}}\}$ \cite{Brun}. To
preserve the trace of the quantum operation these operators satisfy
\begin{equation}
\sum_{n=1}^{N}A_{n}A_{n}^{\dag }=I.  \label{kraus}
\end{equation}%
Let us take two values of the strength parameter $\kappa $: $%
\kappa _{1}$ and $\kappa _{2}$. The corresponding time step operators $%
U_{1} \equiv U\left( \kappa _{1}\right) $ and $U_{2} \equiv U\left(
\kappa _{2}\right) $ are used to define
\begin{equation}
A_{1}\equiv \sqrt{\alpha }~U_{1},  \label{a1}
\end{equation}%
\begin{equation}
A_{2}\equiv \sqrt{\beta }~U_{2},  \label{a2}
\end{equation}%
as a particular set of Kraus operators, where $\alpha \in
\left[0,1\right]$ and $\beta ={1-\alpha }$ in order to satisfy Eq.
(\ref{kraus}). Then the following map for the time evolution of the
density matrix is proposed
\begin{equation}
\rho (n+1)=\alpha ~U_{1}\rho (n)U_{1}^{\dag }+\beta ~U_{2}\rho
(n)U_{2}^{\dag },  \label{map}
\end{equation}%
where $n$ indicates the time $t=nT$. When $\alpha $ (or $\beta $)
vanishes Eq.(\ref{map}) reduces to the well known evolution of the
usual kicked rotor in quantum resonance. In other cases $\alpha $
(or $\beta$) may be thought as the probability per time-step to
apply the operator $U_{1}$ (or $U_{2}$) to the density matrix.

We shall study the previous map in the case when the operators
$U_{1}$ and $U_{2}$ commute ($\left[ U_{1},U_{2}\right] =0$), as is
the case in the primary resonances, in what follows we use the
principal resonance for simplicity. In this case it is easy to
prove, using mathematical induction, that the solution of the map
Eq.(\ref{map}) is
\begin{equation}
\rho (n)=\sum_{j=0}^{n}{\binom{n}{j}}\alpha ^{n-j}{\beta ^{j}}%
~U_{1}^{n-j}~U_{2}^{j}~\rho (0)~U_{2}^{j\dag }~U_{1}^{\left( n-j\right) \dag
},  \label{rhon}
\end{equation}%
where ${\binom{n}{j}}={\frac{n!}{j!(n-j)!}}$. It is important to
point out that Eq.(\ref{rhon}) is a generic solution of
Eq.(\ref{map}) for any couple of unitary operators that commute.
This means that the solution of the map is independent of the
details of the model.

\begin{figure}[th]
\begin{center}
\includegraphics[scale=0.38]{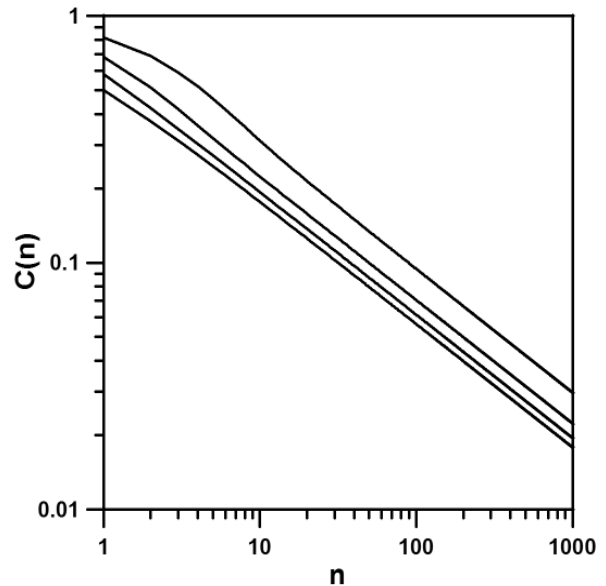}
\end{center}
\caption{The coherence $C(n)$ as a function of the dimensionless
time $n$ in log-log scales for $\Delta \kappa =1000$. The coherence
was calculated, from top to bottom, for $\protect\alpha =0.1$, $\protect\alpha =0.2$, $\protect\alpha %
=0.3$ and $\protect\alpha =0.5$. The straight stretches with slopes
$-0.5$ show a power law behavior
$C(n)\sim\frac{1}{\protect\sqrt{n}}$}
\label{trace1}
\end{figure}
The probability for the angular momentum value $\ell $ at time $n$
is $\texttt{P}(\ell ,n)\equiv\left\langle l\right\vert \rho
(n)\left\vert l\right\rangle $. We shall calculate this probability
for the first principal resonance. The matrix elements of $U_{1}$
and $U_{2}$ are expressed as $\left\langle l\right\vert
U_{1}\left\vert j\right\rangle =i^{-(j-\ell )}\,J_{j-\ell }(\kappa
_{1})$ and $\left\langle l\right\vert U_{2}\left\vert j\right\rangle
=$ $i^{-(j-\ell )}\,J_{j-\ell }(\kappa _{2})$. Then using the above
equation Eq.(\ref{rhon}) with the initial condition $\rho
(0)=\left\vert 0\right\rangle \left\langle 0\right\vert $ the
probability is
\begin{equation}
\texttt{P}(\ell ,n)=\sum_{j=0}^{n}{\binom{n}{j}}\alpha ^{n-j}{\beta ^{j}}%
~\left\langle l\right\vert U(r_{nj})\left\vert 0\right\rangle \left\langle
0\right\vert U^{\dag }(r_{nj})\left\vert l\right\rangle ,  \label{proba}
\end{equation}%
where $r_{nj}=(n-j)\kappa _{1}+j\kappa _{2}$ and $\left\langle
l\right\vert U(r_{nj})\left\vert 0\right\rangle =$
$i^{-l}\,J_{l}(r_{nj})$. The moments of the angular momentum  are
\begin{equation}
\left\langle \ell^{m}(n)\right\rangle =\sum_{\ell =-\infty }^{\ell
=\infty }\ell ^{m}\texttt{P}(\ell ,n).  \label{momentos}
\end{equation}%
We want to study the time behavior of the variance
$\sigma^2=\left\langle \ell ^{2}\right\rangle-{\left\langle
\ell\right\rangle}^2$. The first moment vanishes due to the symmetry
of the initial condition $\rho (0)$. Using the properties of the
Bessel functions, the following value for the variance is obtained
\begin{equation}
{\sigma}^2 (n)=\frac{1}{4}\left[{\left( \alpha \kappa _{1}+{\beta
}\kappa
_{2}\right) ^{2}n^{2}+\left( \kappa _{1}-\kappa _{2}\right) ^{2}\alpha {%
\beta n}}\right].  \label{sigma}
\end{equation}%
In the case when $\kappa _{1}=\kappa _{2}$ the system reduces to the
usual kicked rotor in resonance and its variance has the well known
ballistic behavior characteristic of this case \cite{Izrailev}. When
$\kappa _{1}\neq \kappa _{2}$ the coherence of the system is lost,
as it is shown below, because the probabilistic map is effectively
working. Eq.(\ref{sigma}) leads us to some interesting results. It
shows that the ballistic behavior is maintained with this
decoherence; but additionally there appears the diffusive term
$\left( \kappa _{1}-\kappa _{2}\right) ^{2}\alpha {\beta }n$. In
particular if the parameters verify $\alpha \kappa _{1}+{\beta
}\kappa _{2}=0$, the behavior of the variance is totally diffusive
as in the classical random walk. Then we can conclude that this
decoherence always affects the behavior of the variance but, in
general, does not break its ballistic growth.

The degree of coherence of the system can be measured by several
means. We choose the following:
\begin{equation}
C(n)\equiv Tr\{\rho ^{2}(n)\}=\sum_{l=0}{\left\langle l\right\vert
\rho^{2}(n)\left\vert l\right\rangle }.  \label{coherence}
\end{equation}%
Substituting Eq.(\ref{rhon}) in the above equation, and using the
properties of the Bessel function the equation for the coherence is
obtained
\begin{equation}
C(n)=\sum_{j=0}^{n}\sum_{i=0}^{n}{\binom{n}{j}\binom{n}{i}}\alpha ^{n-j}{%
\beta ^{j}}\alpha ^{n-i}{\beta ^{i}}J_{0}^{2}\left( \Delta \kappa
_{ij}\right) ,  \label{cohe2}
\end{equation}%
where $\Delta\kappa _{ij}=\left( i-j\right)\Delta \kappa$, with
$\Delta\kappa=\kappa_{1}-\kappa_{2}$. From Eq.(\ref{cohe2}) is easy
to prove that $C(0)=1$ and $C(n)<1$ for $n>0$, but in general this
equation will be difficult to reduce to a more simple expression.
However we can get some additional information when $\Delta \kappa $
is very large. In this case $J_{0}^{2}\left( \Delta \kappa
_{ij}\right) $ goes to zero, except when $i=j$ because $J_{0}\left(
0\right) =1$. Then in this limit Eq.(\ref{cohe2}) reduces to
\begin{equation}
C(n)\simeq \sum_{j=0}^{n}{\binom{n}{j}}^{2}\alpha ^{2(n-j)}{\beta
^{2j}}. \label{cohe3}
\end{equation}%
Here we observe the interesting result that the coherence is
independent of the strength parameters of the system if $\Delta
\kappa $ is sufficiently large.

We made numerical studies of the long-time behavior for
Eq.(\ref{cohe2}) as a function of the parameter $\alpha $. In
Fig.~\ref{trace1} the function $C(n)$ is plotted for large $\Delta
\kappa $ and different values of $\alpha $. This figure shows a
power law decay for the coherence $C(n)\sim\frac{1}{\sqrt{n}}$
independently of the value of $\alpha$. The same results were
obtained using Eq.(\ref{cohe3}). Therefore in Eq.(\ref{cohe3}) we
may choose a
particular value of $\alpha $ to calculate its long-time decay. Taking $%
\alpha =1/2$ and using the sums of the binomial coefficients we
obtain
\begin{equation}
C(n)\simeq {\binom{2n}{n}}\left( \frac{1}{2}\right) ^{2n}.  \label{cohe4}
\end{equation}%
For large $n$ is possible to use the Stirling formula to obtain
analytically the following expression for the coherence
\begin{equation}
C(n)\simeq \frac{1}{\sqrt{\pi {n}}},  \label{co}
\end{equation}
confirming our numerical result.

\begin{figure}[th]
\begin{center}
\includegraphics[scale=0.38]{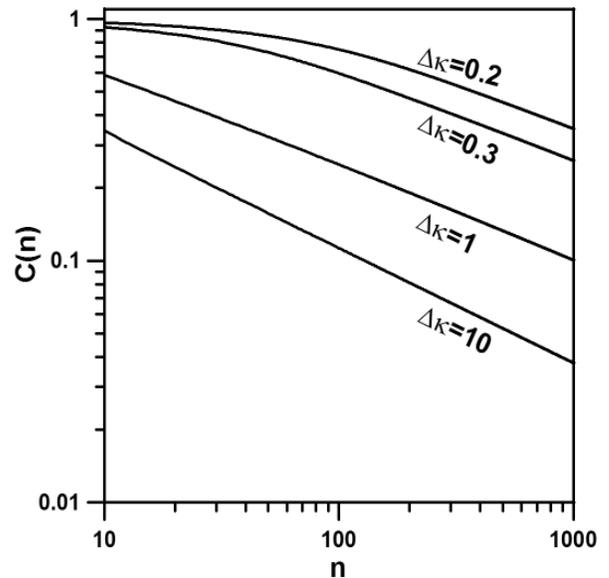}
\end{center}
\caption{The coherence $C(n)$ as a function of the dimensionless
time $n$ for $\alpha=0.1$ in log-log scales. For large $n$ the curve
satisfy a power law $C(n)\sim {{n}^{-\gamma}}$. The parameters of
the curves, from top to bottom, are (1) $\Delta \kappa =0.2$ and
$\gamma=0.36$, (2) $\Delta \kappa =0.3$ and $\gamma=0.37$, (3)
$\Delta \kappa =1$ and $\gamma=0.4$, (4) $\Delta \kappa =10$ and
$\gamma=0.47$} \label{trace3}
\end{figure}

We also studied the coherence for several smaller values of $\Delta
\kappa $. We obtained that for each fixed value of $\Delta \kappa $
the coherence always decays as the power law $C(n)\sim{n^{-\gamma}}$
with $\gamma>0$ (see Fig. ~\ref{trace3}). Additionally we observed
that the exponent $\gamma$ is always independent of $\alpha$ like in
Fig.~\ref{trace1}, therefore $\gamma$ only depends on $\Delta
\kappa$. We observe that the exponent $\gamma$ grows with $%
\Delta \kappa $ with its values in the interval $\left[ 0.3,0.5%
\right] $. We can conclude that the qualitative behavior of
Eq.(\ref{cohe2}) and Eq.(\ref{cohe3}) are the same for all values of
$\Delta \kappa $.

\begin{figure}[th]
\begin{center}
\includegraphics[scale=0.38]{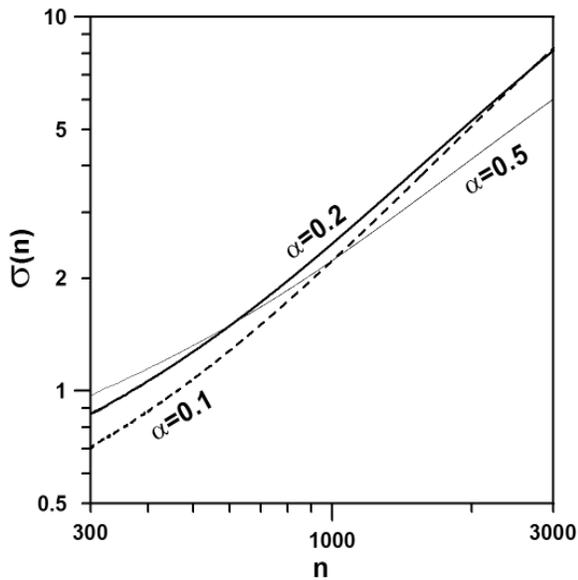}
\end{center}
\caption{The standard deviation $\sigma(n)$ as a function of the
dimensionless time $n$. The parameters are $\protect\kappa
_{1}=0.1$, $\protect\kappa _{2}=0.2$, $p/q=1/3$ in all cases. Dashed
line $\protect\alpha =0.1$
($c=1.2$), thick line $\protect\alpha =0.2$ ($c=1.$), thin line $\protect\alpha %
=0.5$ ($c=0.9$). } \label{trace4}
\end{figure}
Now we inquire the incidence of decoherence on the secondary
resonances. In this case the commutativity between the evolution
operators $U_{1}$ and $U_{2}$ is lost and the expressions for the
variance and the density matrix become very cumbersome. Then we
study the decoherence numerically using Eq.(\ref{map}) for several
values of the parameters $\kappa$ and $\alpha$.

In Fig.~\ref{trace4} the standard deviation $\sigma$ is presented,
for fixed values of $\kappa_1$ and $\kappa_2$ and for different
values of $\alpha$. It is seen that $\sigma$ has power-law decay
with an exponent $c$ that depends on $\alpha$. This parametric
dependence is very different from that given by Eq.(\ref{sigma}) in
the primary resonances where $c=1$. The values of $c$ were adjusted
for the last thousand values of $n$ and we found that they are near
$c=1$. For other values of $\kappa_1$, $\kappa_2$ and $\alpha$ the
exponent $c$ varies between $0,4$ and $1.2$. Considering all the
cases studied we conclude that the exponent $c$ does not show a
clear rule of dependence with the parameters.

The numerical study of the coherence $C(n)$, for the same range of
values of the parameters as for $\sigma$, showed that its time decay
is better approximated by an exponential than by a power law,
\emph{i.e.} $C(n)\sim \exp{(-\delta{n})}$ with $\delta>0$. Therefore
the coherence of the system in the secondary resonances is lost
faster than in the primary ones.

\section{Discussion and conclusion}

Decoherence in quantum systems as QKR or QW has been extensively
studied. Analytical and numerical results
\cite{Kendon1,alejo4,alejo5,Brun} on the effect of different kinds
of noise have shown that quantum properties are highly sensitive to
random events. In particular the linear increase of the standard
deviation $\sigma(t)\sim t$ can be eventually substituted by a
diffusive behavior $\sigma(t)\sim t^{1/2}$ as in the classical
random walk.

The linear increase of the standard deviation of the QKR in
resonance is usually accepted as a direct consequence of the quantum
coherence, in other words, a consequence of the unitary evolution.
This work shows explicitly that unitary decoherence does not break
the temporal linear increase of $\sigma$.

The absence of diffusive behavior in presence of decoherence has
already been shown in our previous works
\cite{alejo1,alejo2,alejo3}. There we have studied the QKR subject
to different types of noise with a L\'evy waiting time distribution
and we found that the system has a sub-ballistic wave function
spreading and its standard deviation has a power-law tail. However
in that opportunity the coherence had not been studied.

Here we have considered a new type of decoherence in the QKR as an
unitary map acting on the density matrix. We obtain an analytical
expression for the density matrix when the Kraus operators commute.
We prove that the decoherence affects the variance but its ballistic
growth persists in spite of an additional linear term. Therefore
asymptotically the linear behavior of the standard deviation is not
suppressed by the noise. On the other hand the coherence $C(n)$ has
a power-law decay for all values of the parameters. We want to
underline that the density matrix Eq.(\ref{rhon}), solution of
Eq.(\ref{map}) only depends on the commutativity of the unitary
operators $U_1$, $U_2$ and it is independent of their detail. This
allows to extend the use of this expression for other quantum models
such as the QW. In  previous works
\cite{auyuanet,alejo1,alejo2,alejo3} we have established a
parallelism between the QKR in resonance with the discrete QW. Then
the type of treatment presented in this paper could be applied to
the QW.

When the Kraus operators do not commute we have not usable
analytical expressions, it is necessary to make numerical studies.
We establish that: a) the standard deviation has no simple
dependence with the parameters of the system, b) the standard
deviation has (in the long-time limit) a continuous range of
behaviors from diffusive to ballistic and c) the coherence $C(n)$
shows a exponential law decay.

We can conclude that the effect of decoherence of the type studied
in this work does not necessarily transform our quantum system into
a dissipative system such as a Markov process. In more general
terms, the mere presence of noise does not assure the passage from
the quantum to the classical world.

We acknowledge the support from PEDECIBA, PDT S/C/IF/54/5, ANII and
thank V. Micenmacher for his comments and stimulating discussions.


\begin{thebibliography}{99}

\bibitem{cohen} C. Cohen-Tannoudji, Rev. Mod. Phys. \textbf{70}, 707 (1998).

\bibitem{Chuang} M. Nielsen, I. Chuang, Quantum Computation and Quantum
Information, 2000. Cambridge University Press, Cambridge.

\bibitem{CCI79} G. Casati, B.V. Chirikov, F.M. Izrailev and J. Ford, \textit{Lect. Notes Phys.}
\textbf{93}, 334 (1979).

\bibitem{Izrailev} F. M. Izrailev, Phys. Rep. \textbf{196, } 299
(1990).

\bibitem{Moore} F.L. Moore, J.C. Robinson, C. Bharucha, P.E. Williams and
M.G. Raizen, Phys. Rev. Lett. \textbf{73}, 2974 (1994).

\bibitem{Kanem} J.F. Kanem, S. Maneshi, M. Partlow, M. Spanner and A.
M. Steinberg, Phys. Rev. Lett. \textbf{98}, 083004 (2007).

\bibitem{Ott} E. Ott, T. M. Antonsen, and J. D. Hanson, Phys. Rev. Lett. \textbf{53}, 2187 (1984).

\bibitem{Dittrich1} T. Dittrich and R. Graham, Z. Phys. B, \textbf{62}, 515 (1986).

\bibitem{Dittrich2} T. Dittrich and R. Graham, Europhys. Lett. \textbf{7}, 287 (1988).

\bibitem{Ammann} H. Ammann, R. Gray, I. Shvarchuck, and N. Christensen,
 Phys. Rev. Lett. \textbf{80}, 4111 (1998).

\bibitem{alejo1} A. Romanelli, R. Siri, V. Micenmacher.
Phys. Rev. E \textbf{76}, 037202 (2007).

\bibitem{alejo2} A. Romanelli.
Phys. Rev. E \textbf{78}, 056209 (2008).

\bibitem{alejo3} A. Romanelli, A. Auyuanet, R. Siri and V. Micenmacher.
Phys. Lett. A, \textbf{365}, 200 (2007).

\bibitem{Schomerus1} H. Schomerus and E. Lutz, Phys. Rev. Lett. \textbf{98}, 260401 (2007).

\bibitem{Schomerus2} H. Schomerus and E. Lutz, Phys. Rev. A \textbf{77}, 062113 (2008).

\bibitem{Casati0} G. Casati, G. Mantica, and D. L. Shepelyansky, Phys. Rev. E \textbf{63},
066217 (2001).

\bibitem{Kraus} K. Kraus, \textit{States, effects and operations: fundamental notions of quantum
theory}, 1983. Springer-Verlag, Berlin.

\bibitem{Brun} T. A. Brun, H. A. Carteret and A. Ambainis, Phys. Rev. A \textbf{67},
032304 (2003).

\bibitem{Kendon1} V. Kendon, Math. Struc. Comp. Sci. \textbf{17}, 1169 (2007).

\bibitem{alejo4} A. Romanelli, R. Siri, G. Abal, A. Auyuanet and R. Donangelo.
Physica A, \textbf{347}, 137 (2004);

\bibitem{alejo5} G. Abal, R. Donangelo, A. Romanelli, A. C. Sicardi
Schifino and R. Siri. Phys. Rev. E \textbf{65}, 046236 (2002).

\bibitem{auyuanet} A. Romanelli, A. Auyuanet, R. Siri, G. Abal and R. Donangelo.
Physica A, \textbf{352,} 409 (2005)



\end{thebibliography}
\end{document}